\documentclass[runningheads]{llncs}
\usepackage[utf8]{inputenc}
\usepackage{threeparttable}

\usepackage{soul}
\usepackage{xcolor} %
\usepackage{filecontents}

\usepackage{dialogue}
\usepackage{tabularx}
\usepackage{float}
\usepackage{amsmath}
\usepackage{subcaption} 
\usepackage[T1]{fontenc}
\usepackage{graphicx}
\usepackage{multirow}
\usepackage{enumitem}
\usepackage{xcolor}
\usepackage[hidelinks]{hyperref}
\usepackage{textgreek}

\usepackage{tabularx} %
\usepackage{booktabs} %
\usepackage{makecell} %
\usepackage{tabularx}
\usepackage{booktabs}
\usepackage{makecell}
\usepackage{float} 
\usepackage{multirow}
\usepackage{orcidlink}
\begin{document}
\title{AI-Driven Assessment of Human Tutors: Linking Training Performance to Real-Life Practice
}
\titlerunning{AI-driven Tutor Training and Assessment} %
\author{Danielle R. Thomas\inst{1}\orcidlink{0000-0001-8196-3252}
\and Marie Cynthia Abijuru Kamikazi\inst{1}\orcidlink{0009-0006-6586-9497} 
\and Clara Brandt\inst{1}\orcidlink{0000-0003-2097-3005}
\and Conrad Borchers\inst{2}\orcidlink{0000-0003-3437-8979} 
\and Kenneth R. Koedinger\inst{1}\orcidlink{0000-0002-5850-4768}
}

\authorrunning{D. R. Thomas et al.}
\institute{
Carnegie Mellon University, Pittsburgh, PA, USA\\
\email{\{drthomas,koedinger\}@cmu.edu}\\
\email{\{mabijuru,clarabra\}@andrew.cmu.edu}
\and
Vanderbilt University, Nashville, TN, USA\\
\email{c.borchers@vanderbilt.edu}
}

\maketitle 

\begin{abstract}

 There exist numerous tutor training platforms. However, few provide AI-driven training and evaluation for human tutors based on real-life performance. We present an AI-driven system that assesses both open responses during training and authentic real-life tutoring. Unlike platforms that only assess learning through online training or simulations, our system utilizes Generative AI (Gemini-2.5-pro) to analyze transcriptions of authentic tutoring, measuring the transfer of tutor skills to real-life application. Human tutors instructing students remotely in math (N=86) completed six scenario-based lessons, averaging a significant 7.4\% learning gain. Using mixed-effects models across 405 session-to-lesson pairs, we found that training performance significantly predicted real-life transcript scores with an effect size of 0.25 SD. Model comparison (AIC/BIC) indicated averaging open response and multiple choice performance during training predicted real-life tutor performance best, although open responses were comparatively more predictive. Exploratory analysis showed that after training, tutors were significantly more likely to encounter pedagogical opportunities to apply their skills (61.1\% to 68.9\%) and demonstrated higher execution quality within those opportunities (65.5\% to 68.1\%). Interrupted time series analysis suggested that these tutor improvements were part of a gradual trend over time rather than an immediate intervention effect of training. We illustrate an AI-driven method to link tutor training with real-life assessment. In doing so, we contribute open datasets, AI prompts, and scoring rubrics to support transparency and reproducibility.

\end{abstract}

\section{Introduction}
The demonstrated efficacy of tutoring has created a surging demand for effective human tutors \cite{nickow2020impressive}. Consequently, the need for trained human tutors is a priority within the technology-enhanced learning community \cite{aleven2023towards}. To successfully scale tutoring programs, ongoing and intensive human training is essential, particularly for volunteers, college students, and paraprofessionals who may lack pedagogical understanding and formal teaching experience \cite{kraft2021blueprint}. Although existing AI-driven professional development (PD) and workforce training platforms offer training via simulations and AI-generated scenarios, they generally lack the ability to verify whether these skills transfer to real-world practice \cite{thomas2023tutor,thompson2019teacher}. 

The inability to assess the transfer of new tutoring skills to real-life applications creates a disconnect between training environments and authentic tutoring interactions \cite{weersink2019simulation}. A key challenge in assessing transfer is that many pedagogical skills are opportunity-dependent: tutors can only demonstrate a given skill when a relevant moment arises. We refer to these moments as \textit{pedagogical opportunities} and distinguish between the likelihood of encountering such opportunities and the quality of a tutor's move when they occur. Distinguishing both is important when determining whether costly and resource-intensive PD actually improves instructional quality and, therefore, student learning outcomes \cite{demszky2024can}. 
To date, little has been done to establish a robust process for validating the external validity and generalizability of human tutor training across diverse tutoring implementations. Additionally, predictive modeling is needed to reveal whether human tutor performance on open responses in training or multiple-choice questions (MCQs) is a stronger driver of  behavioral transfer to real-life application \cite{thomas2025does}. Accordingly, we examine whether AI-driven assessments of tutor training can serve as a signal of real-life application of pedagogical skills. We aim to answer: \textbf{RQ1 (Learning):} Do tutors demonstrate significant learning gains on specific pedagogical skills (``tutor moves'') within AI-enhanced lessons, as measured by pretest to posttest performance improvements? \textbf{RQ2 (Application \& Prediction):} To what extent does performance in training predict the successful application and quality of those skills in real-life tutoring sessions? \textbf{RQ3 (Predictive Formats):} Does the open response or multiple-choice format provide a stronger predictor of behavioral transfer to real-life tutoring?

We introduce an AI-driven pipeline to link tutor performance in online lessons to real-life tutoring behaviors  using generative AI. We contribute LLM prompts and pipelines to score tutor open responses and evaluate the application of tutor moves in real-life tutoring (Accessible in GitHub \cite{anon_repo_2024}). We provide datasets of human tutor training performance (Accessible in DataShop \cite{BlindDataset_2024}).

\subsection{AI-driven Human Tutor Training and Evaluation System}

\begin{figure}[ht]
    \centering
    \includegraphics[width=0.85\textwidth]{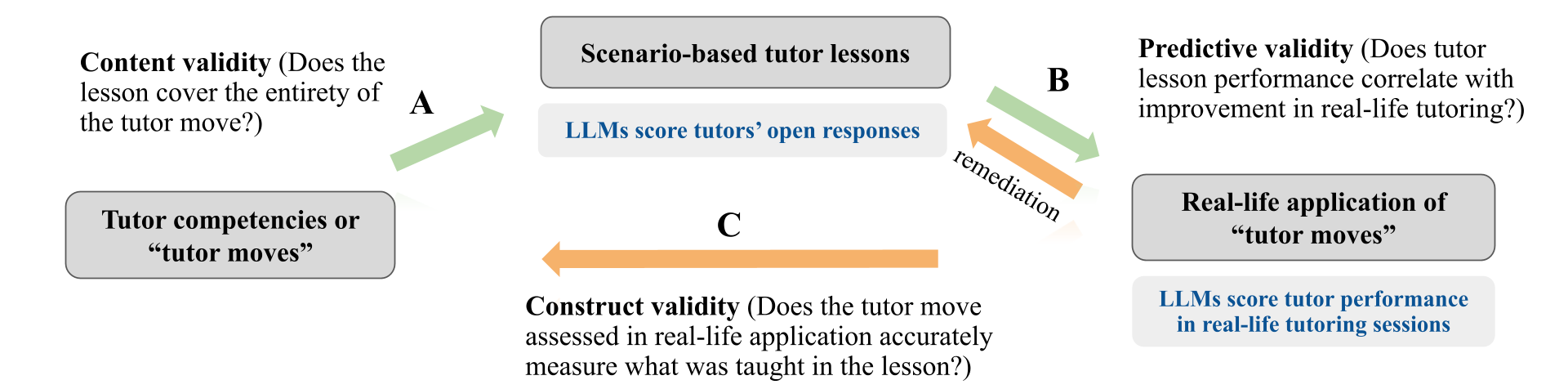}
    \caption{AI-driven tutor training and real-life assessment system.}
    \label{fig:assessment_system}
\end{figure}

Fig.~\ref{fig:assessment_system} illustrates the AI-driven system designed to link tutor training with the assessment of real-life performance. It begins with research-supported tutoring competencies, or {\textbf{{tutor moves}}} and practice via immersive {\textbf{{scenario-based tutor lessons}}}. The process transitions to {\textbf{{real-life application of tutor moves,}}} where LLMs evaluate the quality of tutor moves by determining if opportunities were present, and then evaluating within authentic tutoring transcriptions (e.g., audio, chat). If proficiency is not met, the system triggers a remediation loop of self-reflection or recommends the tutor take the lesson again.

We map three validity types to stages of the system, drawing on standard methodology \cite{trochim2016research}. {\textbf{{Content validity (A)}}} is established via expert judgment to confirm lessons fully cover the ``tutor move'' construct. {\textbf{{Predictive validity (B)}}} assesses whether training success correlates with real-world application as measured by LLM transcript analysis. {\textbf{{Construct validity (C)}}} verifies that AI-driven scoring accurately reflects the intended theoretical pedagogical skills.  

\section{Related Work}
\subsection{Tutor Competencies and Scenario-based Workforce Training}
Tutor competencies, often termed “tutor moves,” are categorized through various theoretical lenses. These include Accountable Talk Moves, which promote equitable classroom discourse \cite{suresh2022talkmoves}; dialogue act theory, which focuses on the structural mechanics of conversational turns \cite{vail2014identifying}; and cognitive science, which distinguishes between knowledge-telling by tutors and the more effective tutor knowledge-building by prompting students \cite{chi2001learning}. Despite these frameworks, empirical evidence linking tutor professional development to real-life application is rare. Scenario-based learning facilitates rapid skill acquisition by embedding instruction in real-world, low-risk contexts \cite{bardach2021power,chine2022development,thompson2019teacher}. Proven effective for teacher and tutor training, this ``learning by doing'' approach ensures skill transfer to authentic sessions \cite{hursen2017investigating,schank2013learning,thomas2023tutor}. Lessons often utilize a modified predict-observe-explain framework grounded in Gibbs’ Reflective Cycle \cite{gibbs1988learning} to drive learning gains of ${\sim}20\%$ \cite{thomas2023tutor,thomas2025llm}. While open-response questions are often assumed to foster deeper learning in these scenarios, recent research suggests that MCQs can be equally effective for tutor advocacy training while offering significantly greater efficiency, reducing the time required for tutors to reach proficiency by 30\% \cite{thomas2025does}.

\subsection{LLMs in Workforce Training and Real-life Application}
{\textbf{Open-Response Scoring using LLMs.}} Recently, LLMs have demonstrated considerable potential in scoring open responses \cite{thomas2025llm}. Previous automated short answer grading, such as those utilizing BERT or Sentence-BERT, often struggled with the linguistic nuances and complex semantics inherent in tutor-student interactions \cite{condor2021automatic}. However, LLMs like GPT-4 have shown the ability to perform nuanced situational reasoning and automated scoring that rivals human-authored support \cite{lee2024applying}. In a study evaluating tutor advocacy training, researchers used GPT-4o and GPT-4-turbo to autograde open responses across scenario-based training  similar to this present work \cite{thomas2025does}. Employing a few-shot prompting strategy, such as providing the model with a limited set of learner-sourced responses scored by  human raters \cite{brown2020language}, the models performed moderately well. While these findings suggest that LLMs offer a viable pathway for scaling the evaluation of complex ``tutor moves'' \cite{lin2024can}, the researchers also identified limitations in highly subjective domains (e.g., advocacy training), where the models showed lower performance compared to simpler pedagogical tasks. These results highlight that while LLMs are effective for low-stakes assessment, further refinement in prompt engineering is required for highly nuanced contexts \cite{thomas2025llm,yun2024enhancing}.

{\textbf{LLM-driven Analysis of Real-life Application of Tutor Moves.}} Little research exists on the use of AI-based methods to evaluate human tutors and teachers using transcription analysis of instructional interaction \cite{borchers2026brief,thomas2025leveraging,wang2024tutor}. One notable example is M-Powering Teachers, an automated NLP-based tool developed by Demszky et al. \cite{demszky2024can} to provide teachers with feedback on their ``uptake'' of student contributions—a high-leverage dialogic teaching practice that makes students feel heard. In a randomized controlled trial (RCT) of M-Powering Teachers involving 1,136 instructors in an online computer science course, researchers evaluated the effectiveness of the tool. They found that the automated feedback improved instructors’ uptake of student contributions by 13\% \cite{demszky2024can}. These results demonstrate the promise of automated feedback tools to complement existing efforts in teachers’ PD, particularly when coaching resources are limited.

\section{Method}
\subsection{Tutor Participants, Study Design, and Data Collection}
\textbf{Tutor Participants}. The study involved 86 undergraduate students from a Mid-Atlantic college in the U.S. who served as paid tutors. They provided semester-long remote tutoring to middle school students in math, and the tutor cohort was culturally and racially diverse. The students, although not research subjects in this present work, were middle school students receiving virtual tutoring during the school day from a dozen schools and five U.S. states. Approximately 80\% of the students came from low-income communities. We prioritized maintaining the privacy and confidentiality of tutors and adhered to all Institutional Review Board (IRB) requirements. 

\begin{figure}[ht]
    \centering
    \includegraphics[width=0.85\textwidth]{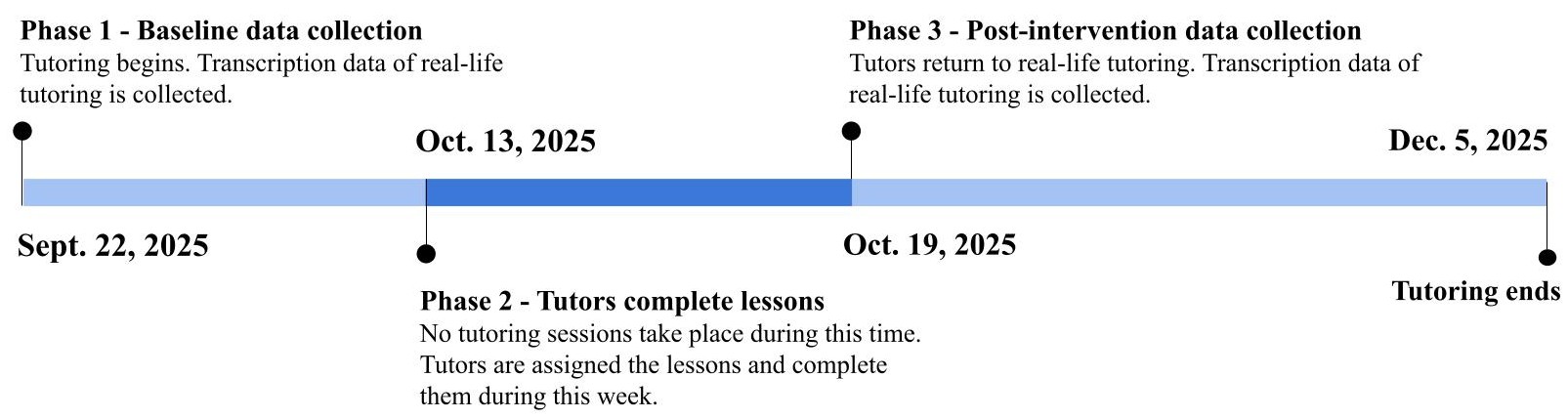}
    \caption{Interrupted-time series research design.}
    \label{fig:ITS}
\end{figure}

\textbf{Study Design}. We utilized a within-subjects interrupted time series design across three phases (Fig.~\ref{fig:ITS}). {\textbf{Phase 1}} established baseline behavior prior to tutors completing online training. In {\textbf{Phase 2,}} tutors paused active tutoring to complete six AI-enhanced scenario-based lessons. In {\textbf{Phase 3,}} tutors resumed real-life tutoring sessions, where audio transcriptions and Zoom chat logs were collected to evaluate the frequency and quality of targeted tutor moves. Tutor lesson log data and transcript analysis allowed for the measurement of construct and  predictive validity by comparing lesson performance and comparing baseline behavior to post-training performance. Data collection consisted of system-generated lesson logs and real-life remote tutoring session data. To generate comprehensive session transcripts, audio from Zoom recordings was processed using OpenAI’s Whisper model and subsequently merged with the corresponding synchronized chat files. Following IRB protocols, all transcripts were deidentified to ensure participant privacy. Tutor lesson data are accessible in DataShop \cite{BlindDataset_2024}.

\subsection{Assessing Tutors’ Scenario-based Lesson Performance}
The lessons focused on six research-supported tutor moves. Three of the lessons covered competencies for which construct validity (see Fig.~\ref{fig:assessment_system}) has been demonstrated previously \cite{thomas2023tutor}: 1) giving effective praise (GIVE\_PRAISE); 2) reacting to errors (REACT\_ERRORS); and 3) determining what students know (DETERMINE\_KNOW) \cite{thomas2023tutor}. Three other lessons aligned with high-quality tutoring strategies identified by Tutor CoPilot, an AI-powered tutoring system that assists human tutors \cite{wang2024tutor}: 4) affirming a correct attempt (AFFIRM\_CORRECT); 5) asking questions to guide thinking (GUIDE\_THINKING); and 6) prompting students to explain (PROMPT\_EXPLAIN). Lessons were co-designed by learning scientists and tutor trainers using authentic, real-life scenarios to ensure face and construct validity (Fig.~\ref{fig:assessment_system}, step A). 

\begin{figure}[ht]
    \centering
    \includegraphics[width=0.9\textwidth]{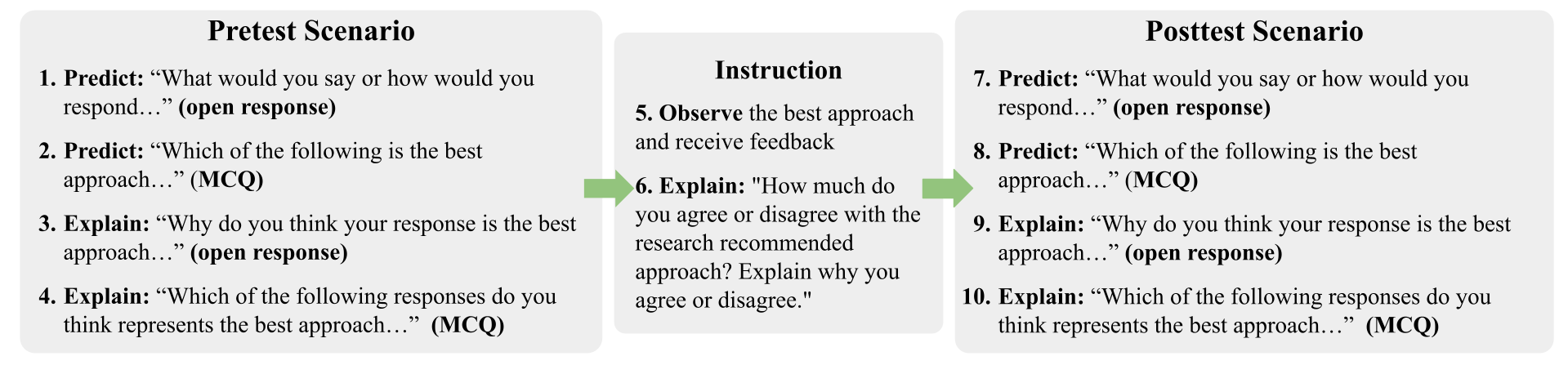}
    \caption{Pretest-posttest instructional design.}
    \label{fig:instructional_design}
\end{figure}

As illustrated in Fig.~\ref{fig:instructional_design}, tutors completed a pretest and had to \textit{predict} and \textit{explain} optimal responses in both open responses and MCQs (1–4). After observing the research-based best approach (5–6), tutors performed a parallel transfer scenario as a posttest (7–10). Pretest and posttest scenarios were counterbalanced. Both tests totaled four points (two MCQs, two open-ended).

Learning gains were determined by subtracting pretest from posttest scores, providing a quantitative measure of skill gains \cite{chine2022development,thomas2023tutor,thomas2025llm}.  While MCQs were automatically scored by the training platform, open responses were evaluated using Gemini-2.5-pro. The AI scoring process was driven by prompts derived directly from expert-validated human scoring rubrics and the process for establishing IRR was documented. Full documentation of the lesson content, human annotation rubrics, and LLM scoring prompts are available in the GitHub \cite{anon_repo_2024}.

\textbf{Inter-rater reliability of open responses}. Inter-rater reliability (IRR) was established by comparing AI scores against a ``human truth'' gold standard. Two expert coders manually graded a subset of 40 responses per lesson (20 \textit{predict} and 20 \textit{explain} responses, see Fig.~\ref{fig:instructional_design}), reconciling disagreements to establish the ground truth. Cohen’s Kappa ($\kappa$) was then calculated to measure the agreement between humans and Gemini-2.5 pro. Table~\ref{tab:irr-scoring} displays reliability results, which were heterogeneous across question types with $\kappa$ values ranging from 0.41 to 1.00 across different tutor moves and question types. 

\begin{table}[htbp]
\centering
\caption{IRR for Open Response Scoring}
\label{tab:irr-scoring}
\fontsize{7pt}{8.5pt}\selectfont %
    \renewcommand{\arraystretch}{1} %

\begin{tabular}{l|cc|cc}
\hline
\textbf{Lesson} & \multicolumn{2}{c|}{\textbf{Human-to-human ($\kappa$)}} & \multicolumn{2}{c}{\textbf{Human-to-LLM ($\kappa$)}} \\
 & \textbf{Predict} & \textbf{Explain} & \textbf{Predict} & \textbf{Explain} \\
\hline
GUIDE\_THINKING & 0.78 & 0.57 & 0.78 & 0.41 \\
AFFIRM\_CORRECT & 1.00  & 0.77 & 1.00  & 1.00 \\
DETERMINE\_KNOW & 0.69 & 0.69 & 0.89 & 0.74 \\
GIVE\_PRAISE   & 0.64 & 1.00  & 0.41 & 0.76 \\
PROMPT\_EXPLAIN & 1.00  & 0.67 & 1.00  & 0.53 \\
REACT\_ERRORS  & 1.00  & 0.89 & 0.69 & 0.67 \\
\hline
\end{tabular}
\end{table}

\subsection{Evaluating Real-life Application of Tutor Moves}
Regarding RQ2 and using a previously described method \cite{thomas2025leveraging}, we utilized Gemini-2.5-pro to identify and evaluate the six tutor moves within session transcripts. This automated assessment pipeline employed a two-stage prompting strategy: an ``opportunity prompt'' identified relevant moments where a tutor move was applicable (0/1), e.g., a student making a math error or a correct attempt; an ``evaluation prompt'' scored the quality of the tutor action following the same evidence-based best practices of the corresponding lesson (0-ineffective, 1-effective). Table~\ref{tab:two-stage-react-error-prompt} shows part of the prompts for the tutor move of reacting to a student error (REACT\_ERRORS). By adapting prompts from validated human coding rubrics, the system provided a scalable measure of skill transfer. Correlation, linear mixed-effects models, and ITS analysis were used to answer RQ2. Full documentation of the lesson content, human annotation rubrics, and full LLM scoring prompts are available in the GitHub \cite{anon_repo_2024}.

\begin{table}[h!]
\centering
\caption{A segment of the two-stage LLM prompt for determining if there was an opportunity for a tutor to react to a student's math error and, if so, evaluating the tutor's response to it (REACT\_ERROR).}

\label{tab:two-stage-react-error-prompt}
\fontsize{7pt}{8pt}\selectfont %
    \renewcommand{\arraystretch}{0.95} %

\begin{tabularx}{\textwidth}{X} %
\toprule
\multicolumn{1}{c}{\textbf{Stage 1: Opportunity Identification Prompt}} \\
\midrule
Your task is to determine whether the following tutoring transcript contains a clear opportunity for a tutor to react to a student's math error. An \textbf{opportunity to react to a student's math error} exists when:
A student makes a mistake while solving a math problem (e.g., The student incorrectly solves \(118 + 18 = 126\)).

\centerline{---}

\textbf{Scoring rules:}
\begin{description}[font=\normalfont, leftmargin=1.5em, style=unboxed, itemsep=1pt]
    \item[\textbf{1:}] If there was an opportunity for a tutor to react to a student's math error.
    \item[\textbf{0:}] If there was NO opportunity for a tutor to react to a student's math error.
\end{description}
If the student and the tutor did not work on any math problem then there was no opportunity. \\

\centerline{---}

\textbf{Constraints:}\\
Focus only on students making math errors, not whether the tutor reacted effectively or ineffectively.
 \\
\midrule[1.2pt] %
\multicolumn{1}{c}{\textbf{Stage 2: Evaluation Prompt (If Opportunity = 1)}} \\
\midrule
Analyze the tutoring audio transcripts below and score the tutor on the following 6 binary (0 or 1) dimensions. For each dimension, return both a binary score (0 or 1) and a short evidence from the transcript under that dimension name. \\

\centerline{---}

\textbf{Rubric Dimensions (Return 0 or 1 for each)} \\

\textbf{1. \texttt{REACT\_ERROR}:}
\begin{description}[font=\normalfont, leftmargin=1.5em, style=unboxed, itemsep=1pt]
    \item[\textbf{0:}] The student made a mistake and the tutor either gave the answer \textbf{OR} pointed out the error directly.
    \item[\textbf{1:}] The tutor responded to a math error by asking the student to explain their thinking \textbf{OR} prompting them to think again.
\end{description} \\
\bottomrule
\end{tabularx}
\end{table}

\textbf{Inter-rater reliability of transcription scoring.} To validate the LLM-based assessment method, two experienced researchers annotated a subset of 10 transcripts across all six tutor moves for the ``opportunity'' and ``evaluation'' prompts shown in Table~\ref{tab:two-stage-react-error-prompt}. In total, researchers evaluated 2,063 lines of transcript dialogue to identify pedagogical opportunities and score tutor execution quality. IRR between the human raters was calculated and shown in Table~\ref{tab:irr-transcript}, and all disagreements were reconciled to establish a ``ground truth'' score. The model was then evaluated on these transcriptions to determine its alignment with expert judgment. Kappa scores for Gemini-2.5-pro compared to ground truth in the lessons ranged from 0.0 (when nearly all transcriptions lacked the given tutor move) to 1.00. However, these scores are rendered less representative of the LLM ability due to the small number of transcripts analyzed, where even minor disagreements can have an outsized impact on the agreement measures.

\begin{table}[htbp]
\centering
\begin{threeparttable}
\caption{Inter-Rater Reliability (IRR) of Transcript Scoring}
\label{tab:irr-transcript}

\fontsize{7pt}{8.5pt}\selectfont
\renewcommand{\arraystretch}{0.95}

\begin{tabular}{lcccc}
\toprule
&
\multicolumn{2}{c}{\textbf{Opportunity Identification}} &
\multicolumn{2}{c}{\textbf{Tutor Evaluation}} \\
\cmidrule(lr){2-3}\cmidrule(lr){4-5}
\textbf{Lesson} &
\makecell{\textbf{Human-Human} \\ \textbf{($\kappa$)}} &
\makecell{\textbf{Human-LLM} \\ \textbf{($\kappa$)}} &
\makecell{\textbf{Human-Human} \\ \textbf{($\kappa$)}} &
\makecell{\textbf{Human-LLM} \\ \textbf{($\kappa$)}} \\
\midrule
DETERMINE\_KNOW & 0.73 & 0.38 & 0.80 & 0.35 \\
GIVE\_PRAISE    & 1.00 & 0.40 & 1.00 & 0.60 \\
REACT\_ERRORS   & 1.00 & 1.00 & 0.00\tnote{*} & 0.00\tnote{*} \\
AFFIRM\_CORRECT & 1.00 & 0.74 & 1.00 & 0.74 \\
GUIDE\_THINKING & 0.74 & 0.52 & 1.00 & 0.38 \\
PROMPT\_EXPLAIN & 0.73 & 0.00\tnote{*} & 0.80 & 0.80 \\
\bottomrule
\multicolumn{5}{l}{\footnotesize *Low $\kappa$ despite high agreement due to data imbalance.}
\end{tabular}

\end{threeparttable}
\end{table}

\subsection{Analytical Methods}
To address \textbf{RQ1}, we measured learning gains from tutoring lessons by calculating the percentage increase from the average pre- to post-test score. We then ran univariate ANOVAs for individual lessons with time as a factor (post vs. pre) to test whether these gains were statistically significant. 

For \textbf{RQ2}, we examined the relationship between tutor training performance and their real-life transcript performance using multiple complementary analyses. First, we computed descriptive Pearson correlations between tutor-level performance on individual lessons and corresponding transcript scores. Second, we estimated linear mixed-effects models predicting transcript scores for each lesson from tutors’ training scores, with random effects to account for repeated measures within tutors. This approach allowed us to formally test whether an overall positive association existed between training performance and transcript outcomes while appropriately modeling within-tutor repeated measures.

To attend to \textbf{RQ3}, we evaluated whether specific types of training performance were more predictive of transcript scores. To do so, we estimated a series of analogous mixed-effects models that replaced overall lesson performance with predictors capturing (a) MCQ performance only, (b) open response performance only, (c) MCQs and \textit{predict} responses, and (d) open responses and \textit{predict} responses. We compared these models to the overall-performance model using AIC and BIC \cite{Baker2025BigDataEducation}. These criteria are used for model selection, where lower values indicate a better-fitting model. They work by balancing a model's goodness-of-fit against its complexity, effectively penalizing the inclusion of unnecessary parameters.  This comparison enabled us to assess whether holistic measures of training performance better predict transcript outcomes than models focused solely on open responses, which may be viewed as more closely approximating authentic tutoring moves in practice. We similarly compared overall performance across pre- and post-test to pre- and post-training performance only.

\section{Results}

\subsection{RQ1: Do Tutors Show Learning Gains in Lessons?}
Table~\ref{tab:tutor_stats} shows tutor lesson performance. The results indicate significant learning gains for the three tutor moves with previously established construct validity REACT\_ERRORS, achieving a 25.2\% gain (\textit{F}=30.10, $p$<$.001$), followed by DETERMINE\_KNOW (14.2\% gain, $p=.006$) and GIVE\_PRAISE (8.1\% gain, $p=.009$). These findings reinforce the efficacy of these established lessons in driving tutor growth in learning. In contrast, the lessons inspired by Wang et al. \cite{wang2024tutor}, AFFIRM\_CORRECT, GUIDE\_THINKING, and PROMPT\_EXPLAIN, did not yield statistically significant learning gains (ranging from 1.4\% to 2.1\%, all $p$>$.05$). This lack of significant growth is likely attributable to a ceiling effect; tutors exhibited exceptionally high baseline performance on these specific moves, with pretest scores already ranging from 0.92 to 0.98 out of 1.0. Despite the flat growth in these newer lessons, the aggregate analysis on all lessons (pooled) revealed a statistically significant learning gain of 7.4\% (\textit{F}=27.50, $p$<$.001$), supporting overall content validity.

\begin{table}[htbp]
    \centering
    \caption{Descriptive statistics and performance of tutors on lessons.}
    \label{tab:tutor_stats}
    \fontsize{7pt}{8.5pt}\selectfont %
    \renewcommand{\arraystretch}{0.95} %
    \begin{tabular}{lcccccc}
        \toprule
        Lesson & Tutors (n) & Pretest (SD) & Posttest (SD) & Score gain & F-statistic & p-value \\
        \midrule
        DETERMINE\_KNOW & 72 & 0.76 (0.29) & 0.87 (0.20) & 14.2\% & 8.194 & 0.006* \\
        GIVE\_PRAISE & 70 & 0.86 (0.18) & 0.93 (0.14) & 8.1\% & 7.305 & 0.009* \\
        REACT\_ERRORS & 66 & 0.70 (0.25) & 0.87 (0.21) & 25.2\% & 30.096 & $<$.001* \\
        \midrule
        AFFIRM\_CORRECT & 75 & 0.98 (0.10) & 0.99 (0.06) & 1.4\% & 1.00 & 0.321 \\
        GUIDE\_THINKING & 74 & 0.95 (0.17) & 0.97 (0.15) & 1.4\% & 0.247 & 0.620 \\
        PROMPT\_EXPLAIN & 76 & 0.92 (0.22) & 0.94 (0.16) & 2.1\% & 0.425 & 0.516 \\
        \midrule
        All lessons (pooled) & 433 & 0.87 (0.23) & 0.93 (0.17) & 7.4\% & 27.495 & $<$.001* \\
        \bottomrule
        \multicolumn{7}{l}{\footnotesize *denotes statistical significance, $p<.05.$} \\
    \end{tabular}
\end{table}

\subsection{RQ2: Does Lesson Performance Predict Real-life Application?}
To examine predictive validity (RQ2), we assessed the relationship between tutors’ lesson performance and their application of the corresponding tutor moves in tutoring sessions. Across tutor–lesson pairs with available transcript data where tutors had the opportunity to demonstrate skills (N = 405), mixed effects models indicated that one SD increase in overall lesson performance was associated with a 0.25 SD increase in transcript scores ($p < .001$).

\begin{figure}[ht]
    \centering
    \includegraphics[width=\textwidth]{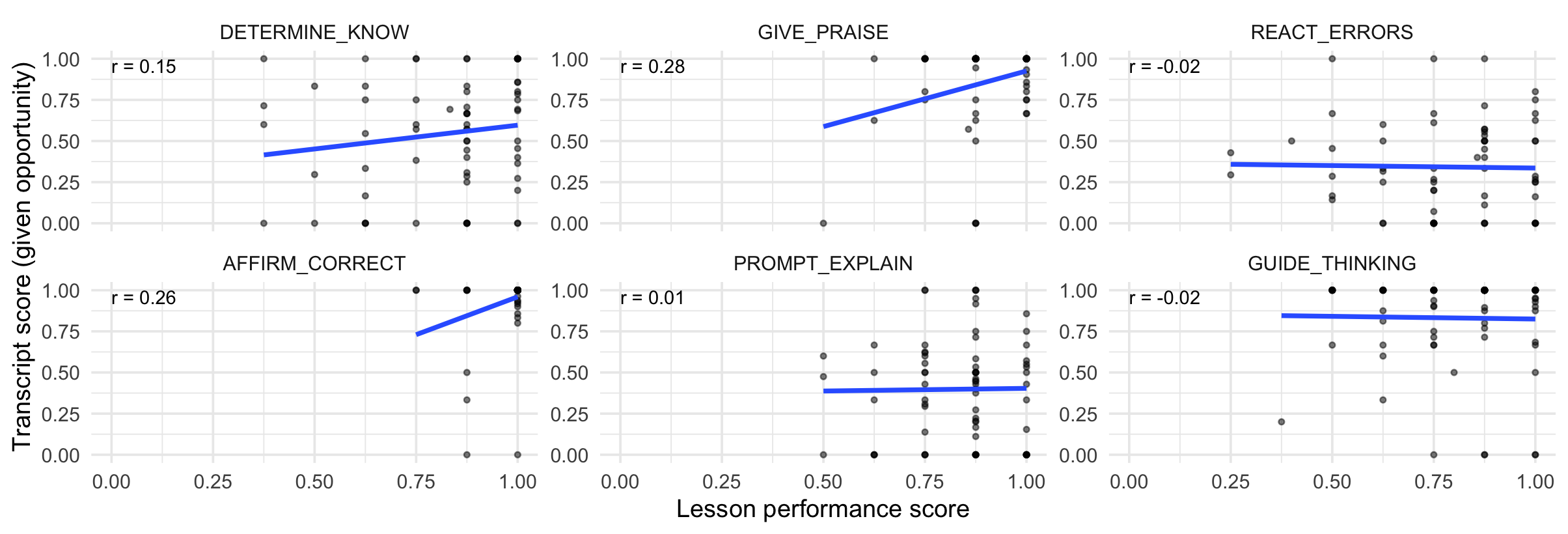}
    \caption{ Linear relationship between lesson performance and assessment of real-life execution, given tutors had the opportunity, by tutor move with correlations.}
    \label{fig:line-graph2}
\end{figure}

Within lessons, tutor-level correlations between tutor training and  real-life performance captured from transcriptions ranged from -0.02 (REACT\_ERRORS) to 0.28 (GIVE\_PRAISE), though only GIVE\_PRAISE ($p = .029$) and AFFIRM\_CORRECT ($p = .030$) were statistically significant while others were not (\emph{p} > .211). Fig.~\ref{fig:line-graph2} displays observed transcript scores as a function of lesson performance, with fitted linear trend lines and Pearson correlation coefficients. Across lessons, it is notable that transcript or lesson performance (or both) often exhibited a ceiling effect with high average scores. Therefore, evidence of transfer from lessons to practice was constrained by distributional limitations.

\subsection{RQ3: Which Question Formats Are Stronger Predictors of Behavioral Transfer to Real-life Tutoring?}
Model comparisons showed that overall training performance provided the best fit to transcript scores (AIC = 1129.7, BIC = 1145.8), outperforming models based on open responses only (AIC = 1130.0, BIC = 1146.1), multiple-choice responses only (AIC = 1150.0, BIC = 1166.0), and models incorporating predicted responses (AICs $\ge$ 1139.2; BICs $\ge$ 1155.2). A model separating coefficients for MCQ and open-response scores did not improve fit (AIC = 1135.1, BIC = 1155.1). Overall, these results indicate that comprehensive measures of training performance yield stronger predictive validity for transcript outcomes than models focused on individual response types. Models estimating tutor training performance via pre- and post-training items did not clearly improve model fit relative to the overall-performance model, with the pre-training model yielding a lower AIC but comparable BIC (AIC = 1118.4, BIC = 1134.4) and the post-training model performing substantially worse (AIC = 1146.0, BIC = 1162.0). Models were fit on different sample sizes and should be interpreted cautiously.

\subsection{Exploration of Tutor Opportunities and Quality Over Time
}
We observed significant improvements over time in both tutors’ opportunities to respond and the quality of their responses. The probability of having an opportunity to execute a target tutoring move was higher post-training than pre-training (post: 68.9\% vs. pre: 61.1\%), a difference that was statistically significant (binomial test, $p < .001$). Conditional on having an opportunity, tutors’ probability of successfully executing the target move also increased from pre- to post-training (65.5\% to 68.1\%; $p = .003$). Together, these results suggest modest but reliable gains in both opportunity and execution quality following training. This strengthens the validity link between scenario-based training performance and real-life transcript performance.

However, we did not gather evidence that changes in tutor quality pre-to-post training were attributable to the training itself. Instead, we observed that tutors generally improved as more weeks passed in the tutoring program. An interrupted time series analysis of average transcript score showed a small but significant linear trend over time ($\beta$ = 0.01, $p = .022$), indicating gradual improvement in tutor quality independent of the training intervention. There was no evidence of an immediate level change at the intervention point (post: p = .246) or a change in slope following training (time × post: $p = .479$). Variation in transcript quality was largely attributable to stable differences between tutors (ICC = .27), with fixed effects explaining little variance (marginal R² = .01).

Tutors' limited post-training performance may reflect constraints in recognizing and capitalizing on opportunities to apply target skills; a competency that is not explicitly trained in the scenario-based intervention, which instead presents tutors with predefined opportunities. A logistic mixed-effects interrupted time series model predicting the likelihood of observing an opportunity to apply a target tutoring skill revealed a significant positive pre-training trend (odds ratio, OR = 1.07 per unit time, $p < .001$), indicating that opportunities were becoming increasingly frequent prior to training. At the intervention point, the estimated level change was negative, suggesting lower immediate odds of opportunity post-training, although this effect was not statistically significant (post OR = 0.63, $p = .111$). Moreover, the post-training slope was significantly attenuated relative to the pre-training trend (time × post OR = 0.99, $p = .028$), indicating a modest slowing in the growth of opportunities following training. Together, these patterns suggest that improvements in tutors’ skill execution may have been constrained by reduced or limited opportunities to deploy those skills in practice (either by recognizing or being presented with relevant tutoring moments), with substantial heterogeneity across tutors and skills remaining (ICC = .35).

Finally, we examined if the linear improvements in tutor performance over time was dependent on them applying skills and learning from these opportunities. We fit a growth curve using a generalized linear mixed-effects model predicting the probability of successful execution conditional on opportunity. An opportunity was defined as a tutoring interaction in which the transcript provided enough context for a specific tutor move to be reasonably demonstrated, as identified by AI-based transcript analysis. The model included cumulative opportunity count, lesson completion (yes vs. no), and their interaction as fixed effects, with a random intercept for tutors. Results indicated no statistically significant effect of opportunity count on successful execution ($\beta$ = 0.06, $p = .39$), suggesting limited evidence of improvement purely through repeated opportunities. The main effect of lesson completion showed a positive but non-significant trend ($\beta$ = 0.35, $p = .098$), and the interaction between opportunity count and lesson completion was also not significant ($\beta$ = -0.07, $p = .36$), indicating that training did not reliably alter the rate of improvement per opportunity.

\section{Discussion \& Future Work}
While the results confirm the system's potential to measure and predict tutoring quality, the updated findings reveal a complex relationship between training interventions and real-world behavior.

\textbf{{Tutor open-response performance is a significant predictor of real-life behavior.}} A primary contribution of this work is the empirical evidence regarding training formats. Our model comparisons demonstrate that performance on open-response items is a stronger predictor of real-life behavioral transfer than MCQ performance. This suggests that the cognitive effort required to generate a pedagogical move, rather than simply recognizing a correct one, more closely mirrors the real-time demands of tutoring. In terms of technology-enhanced learning, this indicates that while MCQs are easier to automate, open responses are essential to validly assess ``readiness'' for authentic tutoring. Aligning with past work, LLM-based grading of open responses has been shown to be an effective method of scaling training containing open responses \cite{thomas2025does}. Still, models that combined open-response and selected-response performance yielded the strongest predictive fit, suggesting that incorporating both assessment formats may be optimal when resources permit. Overall predictive validity, however, was modest (0.25 SD) and varied substantially across lessons. In comparable domains such as medical simulation training, substantially stronger associations have been reported (correlations of approximately 0.6; \cite{weersink2019simulation}). One likely explanation for this discrepancy is that tutors in our context must not only execute high-quality pedagogical responses (e.g., following student errors) but also recognize when such responses are warranted—an ability not explicitly trained in the current scenario-based intervention. This limitation may also help explain lesson-level heterogeneity, as identifying opportunities to react to errors may be more straightforward than recognizing moments that call for prompting explanations or guiding thinking.

\textbf{{Training supports gradual tutor growth.}} Although tutors demonstrated significant learning gains within lessons (7.4\%, $p<.001$) and measurable improvements in real-life quality (65.5\% to 68.1\%, $p=.003$), ITS analysis reveals a lack of a significant immediate performance change at the point of intervention, suggesting that tutoring skill is not acquired through single training events, such as lessons completed at a certain time. Instead, the small but significant linear trend over time ($\beta$ = 0.01, $p=.022$) indicates that growth is a gradual longitudinal process. At the same time, we did not find evidence that opportunities to apply relevant skills explained this linear time trend. An explanation of this finding is that the tutors did not receive feedback on their real-life performance during the semester. Hence, continuous PD with feedback might enhance tutors' gradual improvement over time. One of the most encouraging findings is the significant increase in tutors’ ability to encounter or create pedagogical opportunities (61.1\% pre-training to 68.9\% post-training, $p<.001$). This suggests that the training was effective in helping tutors identify moments where a student’s error or success warranted a specific ``tutor move.'' However, as mentioned, improvements in the quality of execution in those moments rose less strongly. This gap underscores the necessity for continuous feedback. If a tutor performs poorly on executing a ``tutor move'' in real-life, the system can recommend a tutor to complete the corresponding scenario-based lesson. An area of future work is to provide a library of scenarios that can be generated based on the level of difficulty, and is learnersourced from real-life tutoring interactions. The tutor lessons, inspired by \cite{wang2024tutor}, whereby these tutor moves previously demonstrated predictive validity linking to student learning in TutorCoPilot, did not demonstrate construct validity (see Fig.~\ref{fig:assessment_system}). A key reason for this is a content validity limitation regarding over-scaffolding in the newer lessons, particularly in \texttt{PROMPT\_EXPLAIN}. The scenario-based practice questions evaluated tutors on \textit{how} to prompt for self-explanation rather than \textit{when} to do so. Because baseline pre-test scores were high, tutors did not require further instruction on the mechanics of execution. However, despite the lack of active practice on identifying opportunities (\textit{when} to prompt), field-based improvement was indeed demonstrated: post-training transcript analysis showed a significant increase in tutors encountering or creating pedagogical opportunities. This suggests that tutors benefit from passive learning via the explicit instruction, worked examples, and explanations embedded in these lessons, translating this knowledge to real-life practice even when over-scaffolded training questions failed to reflect that growth.

\textbf{Low Kappa scores reflect statistical artifacts of small, imbalanced datasets, not poor model performance.} The human-to-human and human-to-LLM agreement on assessing tutor skills is challenging in practice. This is particularly evident in our transcript scoring results (Table~\ref{tab:irr-transcript}), where several human-to-LLM $\kappa$ scores were low, some between 0.38 and 0.6. As noted, these low scores are often a consequence of two main factors. First, they are impacted by significant data imbalance; for instance, when an opportunity for a specific tutor move is rare, the high agreement on its absence (the majority class) is heavily penalized by the Kappa statistic, which corrects for chance agreement. Second, these IRR measures were calculated on a sample of 10 transcripts, where a single disagreement can have a disproportionate impact on the metric. While 10 transcripts represents a relatively small sample, the annotation was highly granular, requiring manual coding across 2,063 individual lines of dialogue. To illustrate this sensitivity, a change in just one coded label in our data could cause the $\kappa$ to swing from 0.38 (fair agreement) to 0.74 (substantial agreement), highlighting the volatility of the metric with small datasets. In line with recent calls to reframe IRR \cite{thomas2026modernizing}, future work will focus on the continuous refinement of our LLM prompts to better capture the nuance of these pedagogical moves. By automating data collection, we aim to calculate IRR on a much larger and more robust dataset to get a more representative measure of the system's performance.

\textbf{{We demonstrate progress towards an AI-driven system of training tutors and assessing real-life application.}} Referencing Fig. 1, we first established \textbf{content validity (A)} through the expert co-design of our scenario-based lessons, ensuring the lessons fully covered the intended ``tutor move'' constructs. We then established \textbf{predictive validity (B)}. Our mixed-effects model showed that a one SD increase in tutors' lesson performance predicted a 0.25 SD increase in their real-life transcript scores ($p<.001$). Progress was made towards \textbf{construct validity (C)} by developing and validating an AI-driven pipeline capable of identifying pedagogical opportunities and evaluating tutor execution within authentic tutoring transcripts.

\section{Limitations and Conclusion}
Despite the success of Gemini-2.5-pro in these evaluations, the wide range of IRR scores (as noted in Section 3.3) suggests that LLM prompts require continuous auditing. Furthermore, because tutors manually uploaded sessions, a selection bias may exist where more confident tutors submitted more data. Future iterations will focus on automated data collection and the refinement of ``evaluation prompts'' to better capture the nuance of effective moves. Finally, the study acknowledges that the use of LLMs to identify opportunities for tutor moves and evaluate performance is not a perfect process, as indicated by the wide range of IRR (See Table~\ref{tab:irr-transcript} and GitHub \cite{anon_repo_2024}). 

In conclusion, we present a novel AI-driven system designed to train human tutors and assess the real-world application of their skills. The findings demonstrate that while scenario-based training can increase tutors' ability to identify pedagogical opportunities, ongoing, real-life feedback is crucial for improving the quality of their execution. By contributing datasets, AI prompts, and scoring rubrics, this work supports greater transparency and reproducibility, providing a strong foundation for future work on scaling high-quality human tutoring.

\section*{Acknowledgments}
This work was made possible with the support of the Learning Engineering Virtual Institute. The opinions, findings, and conclusions are those of the authors.

\bibliographystyle{splncs04}
\bibliography{main} %

\end{document}